\documentclass[a4paper,twocolumn,secnumarabic,superscriptaddress
,amssymb,amsmath,nobibnotes,aps,prl,floatfix]{revtex4}
\usepackage{docs}%
\usepackage{graphicx}%
\usepackage{bm}%
\usepackage[english]{babel}

\begin{document}

\title{Production of double hypernuclei with high energy antiprotons at PANDA}
\author{Michelangelo Agnello}
\email{michelangelo.agnello@polito.it}
\affiliation{Dipartimento di Fisica, Politecnico di Torino, I-10129 Torino, Italy}
\affiliation{INFN - Sezione di Torino, I-10125 Torino, Italy}
\author{Fabrizio Ferro}
\email{fabrizio.ferro@polito.it}
\affiliation{Dipartimento di Fisica, Politecnico di Torino, I-10129 Torino, Italy}
\author{Felice Iazzi}
\email{felice.iazzi@polito.it}
\affiliation{Dipartimento di Fisica, Politecnico di Torino, I-10129 Torino, Italy}
\affiliation{INFN - Sezione di Torino, I-10125 Torino, Italy}

\date{\today}

\begin{abstract}
The data available in literature, concerning the binding energy of double hypernuclei and their production 
techniques are briefly reviewed. Then, a new technique for producing double hypernuclei with antiprotons in flight 
and measuring their binding energy, proposed for the PANDA experiment at GSI, is investigated. Furthermore, 
preliminary results of the calculations for evaluating the double hypernuclei production and detection rates at the 
antiproton beam intensity foreseen at HESR are reported.
\end{abstract}

\maketitle

\section{Introduction}
The search for Double Hypernuclei (DH) (nuclei with two $\Lambda$'s replacing two non-strange nucleons), 
started in the sixties with the pioneering works of Danysz~{\em et al.} and Prowse~{\em et al.}, who first 
observed $_{\Lambda\Lambda}^{\,10}{\rm Be}$~\cite{Danysz} and 
$_{\Lambda\Lambda}^{\;\;6}{\rm He}$~\cite{Prowse} in emulsions exposed to $K^-$ beams. Both experiments 
measured $B_{\Lambda\Lambda}(_{\Lambda\Lambda}^{\;\;A}{\rm Z})=
B_{\Lambda}(_{\Lambda\Lambda}^{\;\;A}{\rm Z})+B_{\Lambda}(_{\;\;\;\;\Lambda}^{A-1}{\rm Z})$ and $\Delta 
B_{\Lambda\Lambda}(_{\Lambda\Lambda}^{\;\;A}{\rm Z})=B_{\Lambda}(_{\Lambda\Lambda}^{\;\;A}
{\rm Z})-B_{\Lambda}(_{\;\;\;\;\Lambda}^{A-1}{\rm Z})$. The measurement of these quantities is at present the only 
experimental way to collect information about the $\Lambda$-$\Lambda$ interaction and this is the reason of the 
stronger interest for the DH investigation with respect to single hypernuclei. 

Other interesting features of DH are: a) their nuclear structure, with two levels both filled by strange 
hadrons; b) the possibility to explore the existence of the $H$ dibaryon (a double strange system of 6 
quarks predicted several years ago by Jaffe~\cite{Jaffe}); c) the information about the levels 
of the $\Xi^-$-atom, that is formed during the DH production process. In spite of the interest of these topics, 
only few experiments have been devoted to DH so far, mainly because of the difficulty to produce them. The 
traditional way is to produce a double strange $\Xi^-$ particle via the strangeness exchange reaction 
$K^-(p,\Xi^-)K^+$ between a $K^-$ meson and a proton bound in a nucleus, giving $K^+$ and $\Xi^-$; this last, if 
decelerated to rest before decaying, can be captured inside a nucleus in which it interacts with a proton 
eventually releasing two $\Lambda$'s in two hypernuclear levels. The $\Xi^-$ energy may be of order of some 
hundred MeVs, thus the slowing down process takes long time in ordinary matter. This explains why the probability 
of DH formation is so low: in fact, Danysz~\cite{Danysz} and Prowse~\cite{Prowse} observed one DH event each, among 
a $K^-$ number of around a million.

In more recent experiments at BNL-AGS (E885) and KEK (PS E176) about $2\cdot 10^4$ and $800$ stopped 
$\Xi^-$'s respectively were observed, as reviewed by Pochodzalla~\cite{LEAP2003}. Also at the planned 
Japanese Hadron Facility (JHF) the number of produced $\Xi^-$'s is expected to be of order of some 
thousands.

Furthermore the results obtained from experiments in terms of the above mentioned binding energies are 
too much different, clearly outside the experimental errors (see again~\cite{LEAP2003}). 
Thus, while the existence of DH seems ascertained, an increase of statistics in the $\Xi^-$ production is mandatory 
in order to step forward in their understanding.

Even though the traditional double strangeness exchange reaction will be pursued in the next future at JHF, as 
already mentioned, nevertheless the advent of new antiproton facilities, like HESR and JHF (second phase), allows 
the scientific community to explore alternative techniques of $\Xi^-$ production from antiproton beams. A first one 
was proposed by Kilian~\cite{Kilian} who suggested the annihilation of antiprotons at rest to produce $K^-$'s which 
subsequently could exchange strangeness with the residual nucleus, giving a $\Xi^-$ hyperon. This technique 
requires a low energy antiproton machine and could be implemented, for instance, at CERN-AD.

But also in high energy antiproton machines like HESR and JHF, the $\Xi^-$ hyperon may be directly produced in the 
antiproton annihilation reaction together with its antiparticle. The advantages are:
\begin{enumerate}
\item an higher rate with respect to the double reaction involving the annihilation at rest;
\item the presence of the antiparticle, to be used for tagging purposes.
\end{enumerate}

The only disadvantage is the high momentum of the $\Xi^-$'s which need to be strongly decelerated before being 
captured at rest.

In this work we present the first preliminary results of a Monte Carlo simulation based on a simplified 
Intra-Nuclear Cascade Model, and performed to explore the rates of the produced and stopped $\Xi^-$'s and their 
probability to be captured forming a DH, as may be expected in the future PANDA experiment at the HESR-GSI 
facility.

In sec.~\ref{production}, the $\Xi^-$ production techniques are illustrated; in sec.~\ref{section xi 
production}, the details of the antiproton annihilation into the $\Xi^-$'s are discussed; in sec.~\ref{slowing 
down}, the slowing down processes of $\Xi^-$ inside the residual nucleus, in the target and in a second target 
devoted to the DH formation are analysed; in sec.~\ref{results}, the results of our simulation in terms of DH rates 
are reported, and finally in sec.~\ref{conclusions} the conclusions are 
drawn.

\section{Double hypernuclei production with antiprotons}\label{production}
The simplest way for producing double strange nuclear systems is the DH direct formation through $K^{-}K^{+}$ 
reactions; on the contrary, an indirect method is the formation using $\Xi^-$ capture at rest. The latter is known 
to have a higher rate~\cite{Ahn2001}. 

A classical technique, adopted in the E176 experiment at KEK~\cite{Aoki1991,Aoki1991second}, is based on a 
$K^{-}$ kaon beam that induces the quasi-free nuclear reaction $K^{-} +p\rightarrow K^{+} +\Xi^{-}$, thus producing 
a relatively high number of fast $\Xi^{-}$ hyperons. These hyperons subsequently slow down in an emulsion stack 
and some $S=-2$ hypernuclei are formed following the $\Xi^{-}$ capture at rest. In such kind of experiments, 
the emulsion is also the detector, with high enough resolution for tracking the formed hypernuclei and the products 
of their decay as well. A further improvement has been obtained in E373 at KEK~\cite{Ahn2001}, through a $K^{-}$ 
beam (with momentum $\sim 1.66\,{\rm GeV/c}$) impinging on a diamond target in which the same reaction previously 
sketched occurs, though at far higher rate: in the experimental setup, the diamond block was followed by an 
emulsion stack as for E176 and the tracking detector was a fiber-bundle system. In all these experiments, the 
$\Xi^{-}$ hyperon is captured in a light nucleus (i.e. carbon, nitrogen or oxygen) of the emulsion and therefore 
the observed double hypernuclei are $_{\Lambda\Lambda}^{\;\;6}{\rm He}$, $_{\Lambda\Lambda}^{\,10}{\rm Be}$ and 
$_{\Lambda\Lambda}^{\,13}{\rm B}$.

Let us consider now the new technique proposed for the planned PANDA experiment at HESR-GSI; 
it relies on the same method of stopped $\Xi^-$'s, the main difference being that the hyperon production reaction 
is realized through an intense beam of antiprotons on a target of ${^{{\rm A}}{\rm X}^{{\rm N}}_{{\rm Z}}}$ 
nuclei (where ${\rm A}$ is the mass number, ${\rm Z}$ is the atomic number and ${\rm N}$ is the neutron 
number). The two reactions of interest are
\begin{equation}
\bar{p}+{^{{\rm A}}{\rm X}^{\rm{N}}_{\rm{Z}}}
\rightarrow\bar{\Xi}^{0}+\Xi^{-}+{^{\rm{A}-1}\rm{X}^{\rm{N}-1}_{\rm{Z}}}+\rm{mesons}\; ,\label{antiproton-neutron 
reaction}
\end{equation}
and
\begin{equation}
\bar{p}+{^{{\rm A}}{\rm X}^{{\rm N}}_{{\rm Z}}}\rightarrow\bar{\Xi}^{+}+\Xi^{-}+{^{{\rm A}-1}{\rm X'}^{{\rm 
N}}_{{\rm Z}-1}}+{\rm mesons}\; .\label{antiproton-proton reaction}
\end{equation}

In reaction~\ref{antiproton-neutron reaction}, the energetic antiproton annihilates on a neutron 
bound inside the ${\rm X}$ nucleus, while in reaction~\ref{antiproton-proton reaction} it 
annihilates on a proton.

The fast $\Xi^{-}$ slightly slows down in the production target itself, but the main part of its energy-loss 
process occurs in a second separate target, that is also the place in which a statistical collection of double 
hypernuclei is supposed to be formed with some probability. In both 
reactions~\ref{antiproton-neutron reaction} and~\ref{antiproton-proton reaction}, beside the $S=-2$ hyperon 
$\Xi^-$, an anti-hyperon ($\bar{\Xi}^0$ or $\bar{\Xi}^+$) is released too, according to the strangeness and baryon 
conservation laws of strong interactions: this antiparticle can play a crucial role in the experimental detection 
of the whole double hypernuclei formation, as it may be used for trigger purposes.

Before proceeding, we want to point out that a still different technique has also been proposed~\cite{Kilian}, in 
which an antiproton interacts at rest with the ${^{{\rm A}}{\rm X}^{{\rm N}}_{{\rm Z}}}$ nucleus. A possible 
reaction is
\begin{equation}
\bar{p}+{^{{\rm A}}{\rm X}^{{\rm N}}_{{\rm Z}}}\rightarrow
K^{*-}+K^{0}+{^{{\rm A}-1}{\rm X}^{{\rm N}-1}_{{\rm Z}}}+{\rm mesons}\; ,\label{step1}
\end{equation}
where $K^{*-}$ is the $K^{-}$ resonance at $892\,{\rm MeV}$, that subsequently interacts with the residual 
nucleus, giving
\begin{equation}
K^{*-}+{^{{\rm A}-1}{\rm X}^{{\rm N}-1}_{{\rm Z}}}\rightarrow \Xi^{-}+K^{0}+
{^{{\rm A}-2}{\rm X}^{{\rm N}-2}_{{\rm Z}}}+{\rm mesons}\; ,\label{step2}
\end{equation}
where the $\Xi^{-}$ hyperon emerges with a momentum between $250$ and $800\,{\rm MeV/c}$ approximately. In 
comparison with the method based on reactions~\ref{antiproton-neutron reaction} and~\ref{antiproton-proton 
reaction}, this technique has the disadvantage that the $\Xi^{-}$ production occurs in two distinct steps 
(reactions~\ref{step1} and~\ref{step2}), each one characterized by a somewhat low probability.

\section{$\Xi^-$ production}\label{section xi production}
Let us start considering the $\Xi^{-}$ production from reaction~\ref{antiproton-neutron reaction}. A highly 
energetic antiproton $\bar{p}$ interacts with a neutron $n$ bound in a nuclear potential well. In what follows, we 
assume that the neutron be free; indeed, we expect that the total antiproton energy be of order of a few 
GeVs, while the average binding energy of a nucleon is roughly $8.8\,{\rm MeV}$ for the most tightly bound 
nuclei (i.e. $^{56}{\rm Fe}$, $^{58}{\rm Fe}$ and $^{62}{\rm Ni}$).

If we consider $\bar{p}$ momenta below $\pi$ production threshold, the previous discussion allows 
us to consider the reaction as a two-body process, with the manifest advantage that its simple kinematics can be 
treated analytically, starting from the relativistic four-vector energy-momentum conservation; a similar discussion 
also applies to reaction~\ref{antiproton-proton reaction}. Hence, the two reactions simplify as
\begin{equation}
\bar{p}+n\rightarrow\bar{\Xi}^{0}+\Xi^{-}\; ,\label{antiproton-neutron reaction simplified}
\end{equation}
and
\begin{equation}
\bar{p}+p\rightarrow\bar{\Xi}^{+}+\Xi^{-}\; ,\label{antiproton-proton reaction simplified}
\end{equation}
and from now on, we shall refer to them both as Strangeness Creation Reactions, namely SCRs.

The calculated threshold for the two SCRs (neglecting the slight difference between the proton mass and the 
neutron mass) is located at an antiproton momentum of $P^{{\rm thr}}=2.62\,{\rm GeV/c}$. Furthermore, the upper 
limit for the two-body exit channel (i.e. below pion production) is $P^{{\rm thr}}_{\pi}=3\,{\rm GeV/c}$.

\begin{figure}
\begin{center}

\includegraphics[width=\columnwidth]{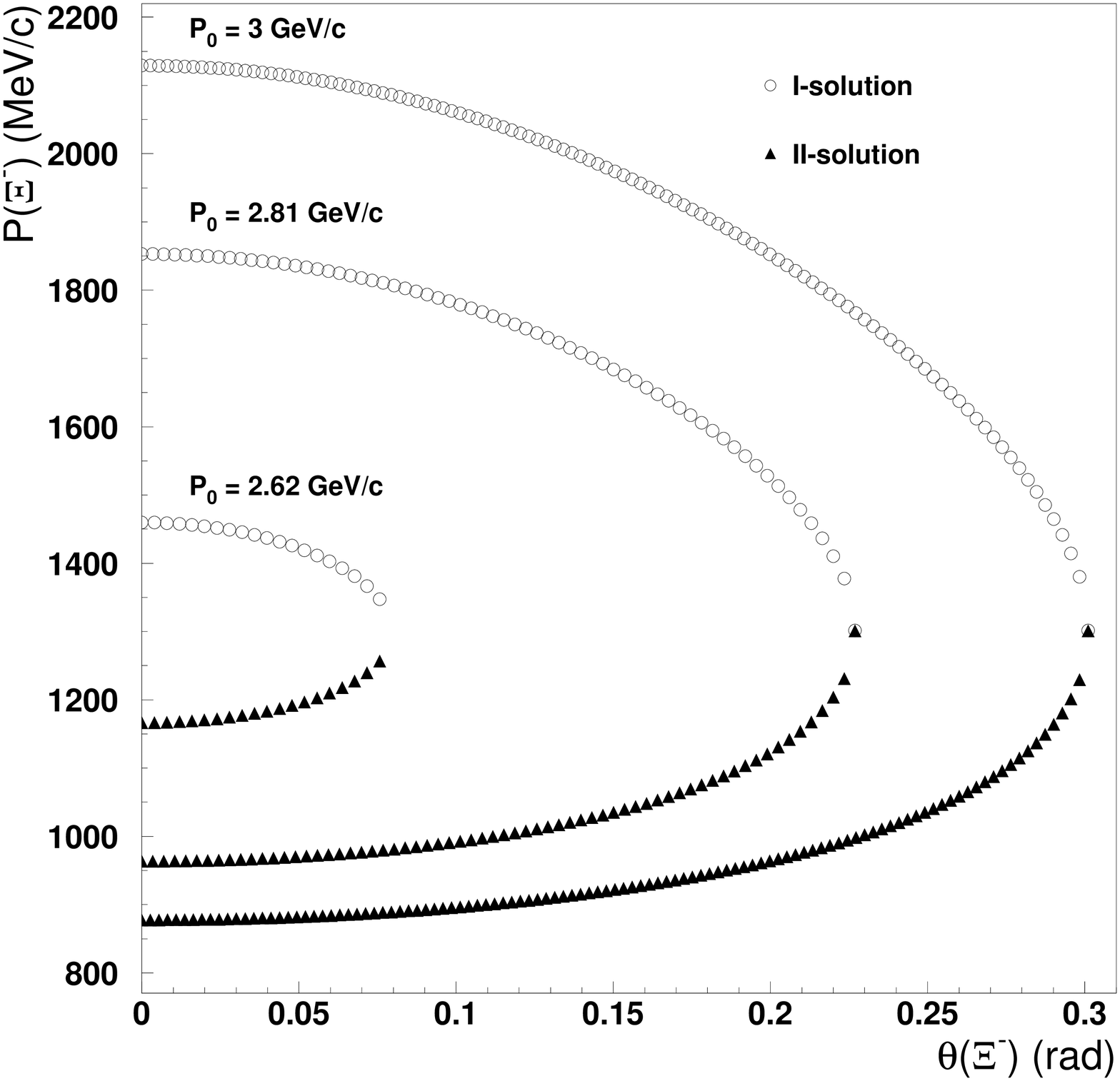}

\caption{Plot of the hyperon momentum $P(\Xi^{-})$ against the exit angle $\theta(\Xi^{-})$ after SCR, 
considering three different values of the antiproton momentum, namely $P_{0}=P^{{\rm thr}}=2.62\,{\rm GeV/c}$, 
$P_{0}=2.81\,{\rm GeV/c}$ and $P_{0}=P^{{\rm thr}}_{\pi}=3\,{\rm GeV/c}$. Both first and second kinematical 
solution are shown together. The curve labelled $P_{0}=3\,{\rm GeV/c}$ has been adopted in our 
calculations.}\label{threeplots}

\end{center}
\end{figure}

Let be $\theta(\Xi^{-})$ the angle between the direction of the incident antiproton $\bar{p}$ and the direction of 
the outgoing $\Xi^{-}$ hyperon, in the laboratory frame of reference. As it is well known from the relativistic 
kinematics of two-body reactions with threshold, the $\theta(\Xi^{-})$ angle ranges between $0$ and a 
maximum, $\theta_{{\rm max}}$, that is a function of the antiproton momentum $P_{0}$. Furthermore, for any given 
value of $\theta(\Xi^{-})$, two different momenta of the $\Xi^{-}$ hyperon are allowed, whose relative probability 
strongly depends on the dynamics of the reaction itself. All these peculiar characteristics are shown in 
fig.~\ref{threeplots} referring to the SCR of eq.~\ref{antiproton-neutron reaction simplified}: the dependence of 
the maximum angle $\theta_{{\rm max}}$ on $P_{0}$ looks evident. At the antiproton momentum 
$P_{0}=3\,{\rm GeV/c}$, we obtain that the maximum angle is $\theta_{{\rm max}}=0.3\,{\rm rad}$.

The argument that led us to the choice of $P_{0}$ relies on the results of the quark-gluon string 
model~\cite{Kaidalov1996} which predicts that the $\bar{p}+p\rightarrow\bar{\Xi}^{+}+\Xi^{-}$ reaction 
cross section $\sigma$ shows a maximum at an antiproton momentum of around $3\,{\rm GeV/c}$; in this case, the 
estimated maximum value is $\sigma_{SCR}\sim 2\,\mu{\rm b}$. Then, in our following calculations, we adopt 
$P_{0}=3\,{\rm GeV/c}$ and we suppose that the same total nuclear cross section, 
$\sigma_{SCR}=2\,\mu{\rm b}$, does hold for both SCRs of eqs.~\ref{antiproton-neutron reaction simplified} 
and~\ref{antiproton-proton reaction simplified}. Lying on these physical hypotheses we obtain that, as far as the 
first (most probable) kinematical solution is concerned, the $\Xi^{-}$ momentum ranges from $2129\,{\rm MeV/c}$ in 
the forward direction, to $1301\,{\rm MeV/c}$ at the limit angle while, considering the second solution, it spans 
from $877\,{\rm MeV/c}$ to about $1301\,{\rm MeV/c}$ respectively, as it can be seen in fig.~\ref{threeplots}.

Because of the high average momentum, the energy-loss process is of central importance in the planned experiment, 
and therefore the evaluation of the $\Xi^{-}$ slowing down is worth while; we have performed it through a 
Monte Carlo simulation, as described in sec.~\ref{slowing down}. As a starting point, we generate $2\cdot 10^5$ 
$\Xi^{-}$ particles produced from SCR for each solution, assuming that reaction~\ref{antiproton-neutron reaction 
simplified} occurs through a S-wave in the centre of mass frame of reference, namely assuming an uniform spherical 
distribution. According to the $3\,{\rm GeV/c}$ curve in fig.~\ref{threeplots}, the momentum spectra of the 
simulated $\Xi^{-}$ hyperons are presented in fig.~\ref{momentum spectra after SCR}, for both kinematical 
solutions.

\begin{figure*}
\includegraphics[width=\textwidth]{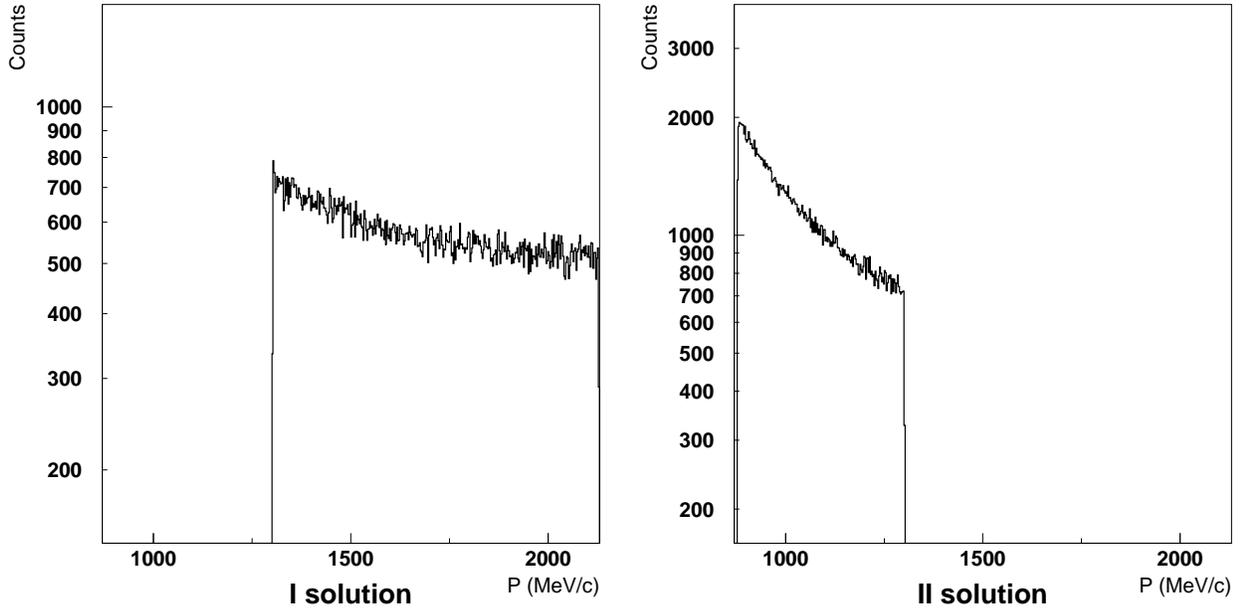}

\caption{Simulated $P(\Xi^{-})$ momentum spectrum after SCR, for both first and second kinematical solution. We 
assume an isotropic angular distribution in the centre of mass frame of reference.}\label{momentum spectra 
after SCR}
\end{figure*}

\begin{figure*}
\includegraphics[width=\textwidth]{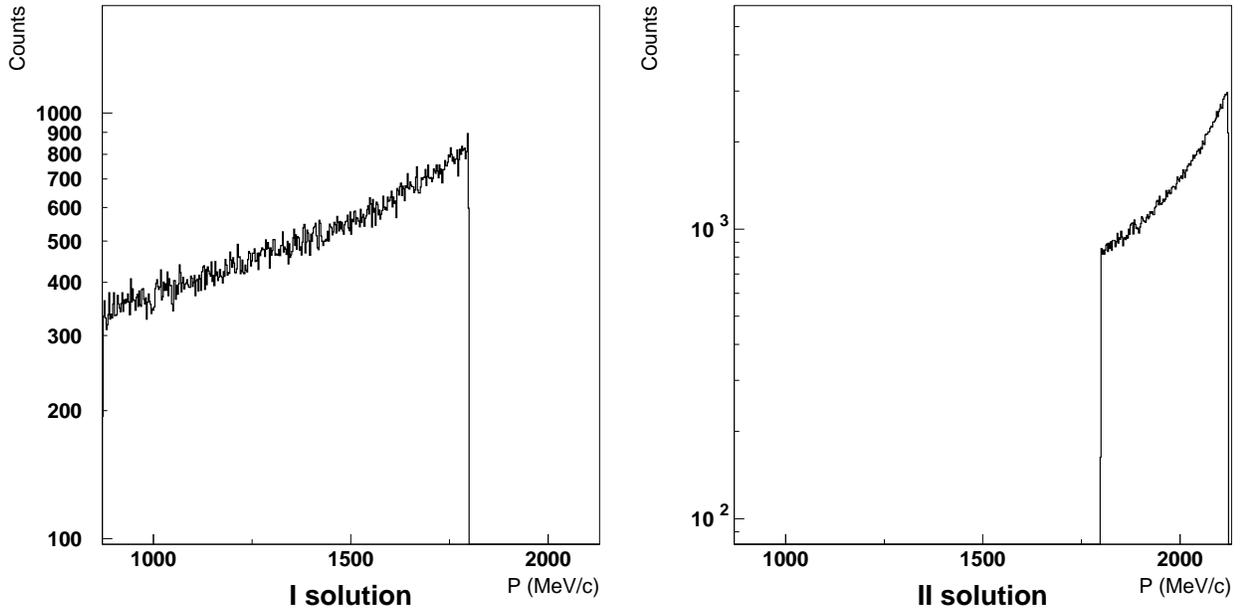}

\caption{Simulated $P(\bar{\Xi}^{0})$ momentum spectrum after SCR, for both first and second kinematical 
solution.}\label{antikaon momentum spectra after SCR}

\end{figure*}

For tagging purposes, the momentum spectrum of the anti-hyperon is of great interest; our computer simulated 
results are reported in fig.~\ref{antikaon momentum spectra after SCR}. The $\bar{\Xi}^{0}$ (or $\bar{\Xi}^{+}$) 
momentum ranges from $871\,{\rm MeV/c}$ to $1798\,{\rm MeV/c}$, with an average emission angle of 
$0.262\,{\rm rad}$, in first solution, and from $1799\,{\rm MeV/c}$ to $2122\,{\rm MeV/c}$ in second solution, 
with an average angle equal to $0.125\,{\rm rad}$. Furthermore, the maximum angle is $0.3\,{\rm rad}$; 
therefore, the tagged anti-hyperon is expected to be released very close to the antiproton beam direction.

\section{$\Xi^{-}$ slowing down}\label{slowing down}
Let us start by noticing that the $\Xi^{-}$ slowing down process consists of two distinct steps:
\begin{enumerate}
\item a sequence of nuclear elastic scattering events with some of the ${\rm A}-1$ nucleons of the residual 
nucleus in which the annihilation has occurred;
\item the energy loss by ionization in the ordinary matter of the two targets.
\end{enumerate}

We shall discuss each step separately.

\subsection{Hyperon-nucleon elastic scattering}\label{scattering}
We assume, on the basis of ref.~\cite{elastic scattering}, that the total elastic cross section for the $\Xi^{-}p$ 
and $\Xi^{-}n$ scattering processes be $\sigma_{{\rm E}}\approx 10\,{\rm mb}$. Moreover, we assume a differential
elastic cross section ${\rm d}\sigma_{{\rm E}}/{\rm d}\Omega\propto\exp(B\cdot t)$, where $t$ is the second
Mandelstam variable, and $B$ is suitably taken as $5\,{\rm GeV}^{-2}$~\cite{Pochodzalla2003}. Furthermore, we 
model the nucleus as a homogeneous sphere of nucleons of radius $R_{{\rm nuc}}=R_{0}\cdot ({\rm A}-1)^{1/3}$ 
($R_{0}\approx 1.35\,{\rm fm}$).

We perform the numerical simulation of this first slowing down step in the framework of an INC-like (Intra
Nuclear Cascade) model~\cite{INC-like}, starting from the distribution of fast hyperons shown in fig.~\ref{momentum 
spectra after SCR}. The basic hypothesis of this model is that the $(A-1)$ residual nucleus does survive for a time 
longer than the time spent by the hyperon during its intra-nuclear path. Furthermore, the scattering exit angle of 
$\Xi^{-}$ after each nuclear scattering is chosen uniformly in the centre of mass frame of reference.

As far as nuclear scattering events are concerned, we treat these processes as instantaneous since the 
calculated average spent time is of order $10^{-22}\,{\rm s}$ (proper time), while the $\Xi^{-}$ mean life, in its 
rest frame of reference, is $\tau_{ML}=1.639\cdot10^{-10}\,{\rm s}$.

We obtain that, from the $2\cdot 10^{5}$ simulated random walks, $9.1\%$ of the hyperons scatter in the residual 
nucleus at least once if we consider the first kinematical solution (the maximum number of consecutive scattering 
events being $3$), and $31\%$ considering the second solution (and the maximum number of scattering events is, 
correspondingly, $8$).

The main physical effects due to hyperon-nucleon elastic scattering are twofold: on one hand, the $P(\Xi^{-})$ 
momentum spectrum is modified with respect to that of fig.~\ref{momentum spectra after SCR}, showing a low momentum 
tail from $1301\,{\rm MeV/c}$ down to $\sim 96.6\,{\rm MeV/c}$ (first solution) and from 
$877\,{\rm MeV/c}$ to $\sim 18.7\,{\rm MeV/c}$ (second solution); on the other hand, the angle 
$\theta_{{\rm nuc}}$ between the exit direction and the antiproton direction ($\theta_{{\rm 
nuc}}\equiv\theta(\Xi^-)$ if the hyperon does not scatter at all), after the nuclear scattering process, spans from 
$0$ up to $1.48\,{\rm rad}$ and $2.38\,{\rm rad}$, respectively, which are much higher than the limit 
angle $\theta_{{\rm max}}$ defined in sec.~\ref{section xi production}. Thus, this first step provides a non 
negligible fraction of hyperons with relatively low momentum and large angles. The next step, involving the passage 
of particles through matter and the ionization energy-loss, further modifies the momentum spectrum, without 
modifying the $\theta_{{\rm nuc}}$ angle. Due to the low probability, the nuclear scattering of the hyperon inside 
the target has been neglected. Thus, the final flight in the experimental apparatus is a simple straight line.

\subsection{Energy-loss by ionization}\label{ionization}
In the study of the energy-loss during the passage of particles through matter, the geometry of the experimental 
setup plays a crucial role, because it determines the effective hyperon path as a function of the exit angle 
$\theta_{{\rm nuc}}$, and thus the slowing down effectiveness.

We adopt here a simple geometry consisting of two distinct targets: the first one, in which the SCR takes
place, is a thin parallelepipedal wire of Gallium or Gold ($4\,{\rm cm}$ long and $5\,\mu{\rm m}\times 
5\,\mu{\rm m}$ square) orthogonal to the beam direction, while the second one is a diamond parallelepiped 
($4\,{\rm cm}\times4\,{\rm cm}$ as transverse dimensions and $2\,{\rm cm}$ thick) with a hole of radius 
$0.25\,{\rm cm}$ in its centre. The two targets are separated by a $1\,{\rm mm}$ vacuum gap.

The physics of our simulation essentially relies on the Bethe-Bloch equation, which expresses 
the mean rate of energy-loss per unit length: we take it from~\cite{particle data} and we report it here for the 
sake of discussion
\begin{eqnarray}
-\frac{{\rm d} E_{K}}{{\rm d} x}&=&
K\rho\frac{Z}{A}\frac{1}{\beta^2}\left[\frac{1}{2}\ln\left(\frac{2m_{{\rm e}}{\rm c}^2\beta^2\gamma^2
T_{{\rm max}}}{I^2}\right)-\beta^2\right.\nonumber\\
& &\left.-\frac{C}{Z}-\frac{\delta}{2}\right]\; 
,\label{Bethe Bloch}
\end{eqnarray}
with
\begin{equation}
T_{{\rm max}}\equiv\frac{2m_{{\rm e}}{\rm c}^2\gamma^2\beta^2}{1+2\gamma\frac{m_{{\rm e}}}{M_{0}}+
\left(\frac{m_{{\rm e}}}{M_{0}}\right)^2}\; ,
\end{equation}
where $E_{K}$ is the hyperon kinetic energy, $\rho$ is the density of target material, 
$K=0.3\,{\rm MeV}\cdot{\rm cm}^2$, $Z$ and $A$ are the atomic number and the mass number of medium, $\beta$ is 
the dimensionless velocity of hyperons, $m_{{\rm e}}$ is the electron mass, $\gamma=(1-\beta^2)^{-1/2}$ is the 
relativistic factor, $I$ is the mean  excitation energy, $M_{0}$ is the hyperon rest mass and ${\rm c}$ is the 
speed of light. The parameters $C$ and $\delta$, usually negligible at high energy, play an important role in our 
similation because the final stage of the slowing down process takes a major fraction of the total time spent by 
the hyperon before being stopped; the value of $\delta$ has been taken from the accurate work of 
Sternheimer~{\em et al.}~\cite{Sternheimer1984} and reads
\begin{equation}
\delta=
\begin{cases}
2X\ln10+C & X\ge X_1\\
2X\ln10+a(X_1-X)^m+C & X_0\le X\le X_1\\
0 & X\le X_0
\end{cases}\; ,
\end{equation}
with $X=\ln\beta\gamma$, and where $C$, $a$, $m$, $X_0$, $X_1$ are taken from ref.~\cite{Sternheimer1984} for the 
diamond target.

The total time $T$ elapsed during the energy-loss process is comparable with the $\Xi^-$ mean life, $\tau_{ML} $: 
thus, a fraction of hyperons decays prior to complete stopping. 

In order to evaluate the number of surviving $\Xi^-$'s, the following procedure has been adopted. Each $\Xi^-$ has 
been assigned a lifetime $\tau$ according to the $\exp(-\tau/\tau_{ML})$ distribution, in the frame in which 
the particle is at rest; at each interval of lost kinetic energy ${\rm d}E_K$, the corresponding time interval 
${\rm d}t$ (in the laboratory frame) spent by $\Xi^-$ to travel a path ${\rm d}x$ has been transformed into the 
rest frame in order to obtain the total proper time elapsed  $T$, and compare it with the proper lifetime $\tau$. 
Thus, eq.~\ref{Bethe Bloch} should be solved provided that the condition
\begin{equation}
T=\int_{0}^{t_{F}}{\rm d} t \sqrt{1-\beta^2}\le\tau\; ,\label{time condition}
\end{equation}
be satisfied, where $t_{F}$ is the final time after complete stopping of the hyperon in the laboratory frame of 
reference.

Eq.~\ref{Bethe Bloch} can be formally expressed through
\begin{equation}
\frac{{\rm d} E_{K}}{{\rm d} x}=-g(E_{K})\; .\label{formal Bethe Bloch}
\end{equation}

Furthermore, the hyperon momentum $P$ depends over $E_{K}$ as
\begin{equation}
P(E_{K})=\sqrt{E_{K}(E_{K}+2M_{0})}\; .\label{momentum-energy relation}
\end{equation}

From eqs.~\ref{time condition}, \ref{formal Bethe Bloch} and~\ref{momentum-energy relation}, we straightforwardly
obtain
\begin{equation}
R=-\int_{E_{K}^{{\rm in}}}^0{\rm d} E_{K}\frac{1}{g(E_{K})}\; ,\label{range}
\end{equation}
\begin{equation}
T=-\frac{M_{0}}{{\rm c}}\int_{E_{K}^{{\rm in}}}^0{\rm d} E_{K}\frac{1}{g(E_{K})P(E_{K})}\le\tau\, ,\label{time}
\end{equation}
where $R$ is the stopping range and $E_{K}^{{\rm in}}$ is the hyperon's initial kinetic energy. Performing a 
numerical integration of eq.~\ref{range}, at each step the corresponding time integral of eq.~\ref{time} is 
calculated and checked whether $T$ is lower or greater than the lifetime $\tau$. In the former case the integration 
proceeds until complete stopping, otherwise the hyperon decays along its path.

\section{Simulation results}\label{results}
The simulation procedure described in the previous section has been applied to the above mentioned geometry (a wire 
as $\Xi^-$ production target followed by a second one as hypernuclear target); furthermore we run the simulation 
twice, considering two different materials (gallium and gold) for the wire, in order to check the effectiveness of 
the atomic weight on hyperon production. For both materials, $2\cdot 10^{5}$ $\Xi^-$'s have been generated in a 
spot of radius $r=10\,\mu{\rm m}$ on the surface of the wire.

The calculated parameters of our simulation are:
\begin{itemize}
\item the fraction $f_{{\rm stop}}$ of hyperons stopped in the second target;
\item the fraction $f_{{\rm dec}}$ of hyperons decayed before stopping;
\item the fraction $f_{{\rm loss}}$ of hyperons lost in the gap between the two targets or in the central hole of 
the second target.
\end{itemize}

Since reaction~\ref{antiproton-neutron reaction} can produce $\Xi^-$'s of two different momenta, both kinematical 
solutions have been simulated separately.

The most relevant parameters for the experiment are the total number $N_{{\rm stop}}$ of stopped and detected 
$\Xi^-$'s per second and the number $N_{\Lambda\Lambda}$ of produced and detected double hypernuclei per second. 
In order to calculate these values from the simulated fractions already defined, one has to make a few reasonable 
assumptions about the luminosity of the machine and the detection efficiency; moreover, other physical parameters 
mentioned below have been roughly estimated just to get a first insight of the rates, though further accurate 
evaluation is needed.

As already discussed in sec.~\ref{section xi production}, we have assumed a total cross section 
$\sigma_{SCR}\approx2\,\mu{\rm b}$ for the SCR reaction between an antiproton and a free nucleon at 
$3\,{\rm GeV/c}$; accordingly, the corresponding reaction cross section of reaction~\ref{antiproton-neutron 
reaction} occurring on a neutron inside a nucleus reads
\begin{equation}
\Sigma_{SCR}\approx\sigma_{SCR}\cdot A^{2/3}\cdot \frac{A-Z}{A} \; ,
\end{equation}
where the surface annihilation scaling has been taken into account~\cite{Iazzi2001}.

Furthermore, we assume a luminosity of the antiproton beam of roughly ${\cal L}\approx 
10^{32}\,{\rm cm}^{-2}{\rm s}^{-1}$, as realistically expected at HESR, and a reconstruction efficiency for the 
whole detector $\varepsilon_{K}\approx 0.5$.

Following refs.~\cite{Pochodzalla2003,conversion}, we also assume that the conversion probability (i.e. the 
probability for the hyperonic $\Xi^{-}$-nucleus conversion into a $\Lambda\Lambda$-hypernucleus) per 
stopped $\Xi^{-}$ be $p_{\Lambda\Lambda}\approx 0.05$, and that the level population fraction be $p_{LP}\approx 
0.1$; the probability of transition per event has been taken as $p_{T}\approx 0.5$ and the $\gamma$-photo peak 
efficiency as $\varepsilon_{\gamma}\approx 0.1$.

Thus, the number of stopped $\Xi^-$'s and of formed $\Lambda\Lambda$-hypernuclei can be calculated through
\begin{equation}
N_{{\rm stop}}={\cal L}\cdot\Sigma_{SCR}\cdot f_{{\rm stop}}\cdot \varepsilon_K\; ,
\end{equation}
and
\begin{equation}
N_{\Lambda\Lambda}=N_{{\rm stop}}\cdot p_{\Lambda\Lambda}\cdot p_T\cdot p_{LP}\cdot \varepsilon_{\gamma}\; .
\end{equation}

The final results are reported in tab.~\ref{table of results} for both gallium and gold wires and both kinematical 
solutions.

\begin{table}[t]

\begin{center}
\caption{Results of the Monte Carlo simulation of the double hypernuclei production, considering two materials 
(gallium and gold) for the hyperon production wire. Each kinematical solution has been investigated separately, 
and the corresponding results are shown. (See text for the physical meaning of the different parameters).}

\vskip 0.1 in

\begin{tabular}{|c||c|c||c|c|} 
\hline
& \multicolumn{2}{c||}{\bf Gallium wire} & \multicolumn{2}{c|}{\bf Gold wire} \\
& I & II & I & II \\
\hline \hline

$f_{{\rm stop}}$ & $5.75\cdot 10^{-4}$ & $1.15\cdot 10^{-2}$ & $1.45\cdot 10^{-3}$ & $2.01\cdot 10^{-2}$\\
\hline
$f_{{\rm dec}}$  & $7.98\cdot 10^{-2}$ & $1.31\cdot 10^{-1}$ & $8.72\cdot 10^{-2}$ & $1.47\cdot 10^{-1}$\\
\hline
$f_{{\rm loss}}$ & $3.76\cdot 10^{-2}$ & $4.44\cdot 10^{-2}$ & $3.73\cdot 10^{-2}$ & $4.39\cdot 10^{-2}$\\
\hline
$N_{{\rm stop}}\;[{\rm s}^{-1}]$     & $0.460$ & $9.224$ & $1.16$ & $16.07$\\
\hline
$N_{\Lambda\Lambda}\;[{\rm s}^{-1}]$ & $1.15\cdot 10^{-4}$ & $2.31\cdot 10^{-3}$ 
                                     & $2.90\cdot 10^{-4}$ & $4.01\cdot 10^{-3}$\\
\hline

\end{tabular}
\label{table of results}
\end{center}
\end{table}

The relative weight of each solution is at present totally unknown as no measurements of the differential 
cross section for reactions~\ref{antiproton-neutron reaction simplified} and~\ref{antiproton-proton reaction 
simplified} exist nor theoretical estimates as well. Hence, the actual value of $N_{\Lambda\Lambda}$ is located
between the two values of first and second solution reported in tab.~\ref{table of results}. Being conservative, 
one may take the value of the first solution only; nevertheless, even in this pessimistic hypothesis, the total 
number of stopped $\Xi^-$'s is around $1.19\cdot 10^6$ per month and the total number of formed double hypernuclei 
is around $300$ per month considering the gallium wire, while we obtain $3\cdot 10^6$ and $750$ respectively when 
considering the gold wire.

On the other hand, if the $\bar{p}$ annihilation occurs through reaction~\ref{antiproton-proton reaction} instead 
of reaction~\ref{antiproton-neutron reaction}, the processes of $\Xi^-$ slowing down and double hypernuclei 
formation are in practice identical to those considered so far: therefore, it is reasonable to expect that the 
total $\Xi^-$ production and double hypernuclei formation should be enhanced by a factor $A/(A-Z)\sim 1.8$.

As a first indication from the values of the stopped $\Xi^-$'s, one can observe that the weight of the nucleus 
strongly influence the hyperon slowing down through nuclear scatterings.

Of course these results depends on the estimates of the machine, detector and physical parameters chosen for our 
simulation, but the values assumed here seem quite realistic indeed. In fact, HESR design's first priority is 
the high luminosity and the PANDA apparatus will be an ensemble of several detectors with good reconstruction 
capabilities.

\section{Conclusions}\label{conclusions}
A preliminary evaluation of the rates for the complete stopping of a $\Xi^-$ hyperon, produced through a high 
energy antiproton interacting with a gallium or a gold nucleus, in a diamond (carbon) target has been performed 
using a Monte Carlo technique based on an INC-like model. The chosen geometry is suitable for the PANDA experiment 
project at the future HESR machine at GSI.

The results, relying on the expected performances of the apparatus, show the feasibility of a large production of 
double hypernuclei: our preliminary estimates, in terms of stopped $\Xi^-$'s and formed double hypernuclei, are 
larger than the previous data existing in literature and also larger than the expectation of other future 
experiments and machines.

A lot of calculations are of course still necessary in order to fully design the geometrical arrangement, the sizes
and the materials of the targets, in order to optimize the production and to meet the beam requirement of the 
machine, and this will be the future work for the next months.



\end{document}